\documentclass[twocolumn]{article}

\usepackage{upgreek}
\usepackage{amsmath}
\usepackage[switch]{lineno} 
\usepackage{graphicx}
\usepackage{siunitx}
\usepackage{bm}
\usepackage{paralist}
\usepackage{dirtytalk}
\usepackage{mathtools}
\usepackage{physics}
\usepackage{hyperref}
\usepackage[thinc]{esdiff}
\usepackage{authblk}
\hypersetup{colorlinks,allcolors=black}
\usepackage[backend=biber, style=nature, isbn=false, eprint=false, url=false, doi=false]{biblatex}

\AtBeginDocument{\RenewCommandCopy\qty\SI}

\begin{document}
\title{Universality in multispecies traffic}

\author[1]{Georg Anagnostopoulos \thanks{Correspondence and requests for materials should be addressed to G.A. (email: georgios.anagnostopoulos@epfl.ch).}}
\author[1]{Nikolas Geroliminis}
\affil[1]{Urban Transport Systems Laboratory, School of Architecture, Civil and Environmental Engineering, EPFL, CH-1015, Lausanne, Switzerland}

\onecolumn
\maketitle

\begin{abstract}
Understanding the nature of traffic heterogeneity is of major importance, given the widespread adoption of micromobility in cities. Based on massive field data and a nonequilibrium model, we demonstrate that heterogeneous, multispecies traffic is a member of an inherently nonequilibrium universality class associated with porous flows, namely directed percolation (DP) in one spatial dimension. Our central finding is that, macroscopically, multispecies traffic behaves like water percolating through a porous medium. This hypothesis remained unresolved for years mainly due to the incompatibility of equilibrium approaches with phenomena that are quite far from equilibrium and the limited resonance of complexity theory in the transportation literature. DP entails the existence of a nontrivial phase transition from a disordered subcritical phase to an ordered supercritical phase that depends on  a temperature-like control parameter and is governed by a universal power-law. Our model explains the large scatter found in experimental data by taking into account the nonlinear, stochastic perturbations present in multispecies traffic configurations due to coupling of a predominantly lane-based host system with a layer of lane-free parasitic flows.
\end{abstract}

\twocolumn

\begin{figure}
\centering
\includegraphics[width=\linewidth]{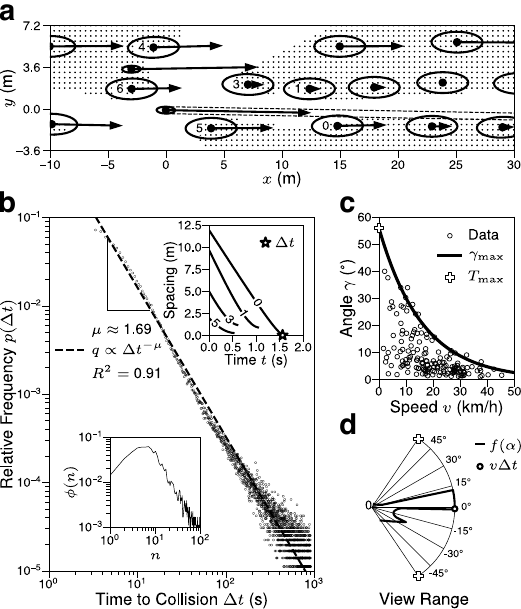}
\caption{Empirical results from the pNEUMA data. (\textbf{\fontfamily{phv}\selectfont
a}) Snapshot of an actual urban arterial road with multispecies traffic. The different vehicle geometries are approximated as inscribed hard elliptical discs. Arrows indicate velocity vectors in m/s. Using the symmetric shadowcasting technique, we perform a visibility analysis in discrete space with a resolution of 45 cm, small enough to also detect microvehicles. (\textbf{\fontfamily{phv}\selectfont
b}) Upper inset: Assuming constant velocities, we compute the minimum time to collision $\Delta t$ between the microvehicle at (0,0) and its visible, converging neighbors $\left\{0,1,3,5\right\}$. Main panel: at the sample level (Supplementary Data 1), the histogram $p(\Delta t)$ has a power-law objective distribution $q\propto\Delta t^{-\mu}$. Lower inset: the same data yields a frequency distribution $\phi(n)$ of the bin counts $n$, which is also heavy-tailed \supercite{thurner2018}. (\textbf{\fontfamily{phv}\selectfont
c}) View range $\gamma_{\max}$ is a function of vehicle speed, with data points showing actual maneuver events (Methods, Supplementary Data 2). The intercept is $T_{\max}$. (\textbf{\fontfamily{phv}\selectfont
d}) Interpretation of the same situation as in the first-panel, but from the agent's spatiotemporal point of view, considering the dynamic porosity $f(\alpha)$.}
\label{fig:fig1}
\end{figure}

Several years ago, it was conjectured that heterogeneous traffic behaves like a porous flow, \say{where smaller vehicles \emph{percolate} to the front of the queue}\supercite{nair2011}, much like water infiltrating a bed of coffee grounds. Surprisingly, the authors of that early paper overlooked the \emph{already established percolation theory} \supercite{stauffer1994, hinrichsen2000}, leaving their conjecture unsubstantiated to this day. Understanding the nature of traffic heterogeneity is evidently of major relevance, particularly given the widespread adoption of \emph{micromobility} \supercite{OHern2020} in cities, where distinct populations of vehicles often share the infrastructure. An example of such a multispecies \supercite{mason1997} system (see Fig. \ref{fig:fig1}a and \href{https://github.com/GAnagno/multispecies-traffic/blob/main/movie-s1.gif}{Supplementary Movie 1}) is thoroughly documented in the pNEUMA \supercite{barmpounakis2020} traffic dataset, which includes half a million trajectories spanning an expansive 1.3 km$^2$ area in the congested central business district of Athens, Greece. 

Empirically motivated by the pNEUMA data and leveraging simulation techniques as a stepping stone, we show that multispecies traffic is a member of the \emph{universality class} associated with porous flows, namely directed percolation (DP) \supercite{hinrichsen2000} in one spatial dimension (1D). DP universality is intimately linked to the idea of \emph{self-organized criticality} (SOC) \supercite{bak1987,stanley1987}, which implies the existence of critical attractor points that are reached by the system spontaneously without meticulous tuning of control parameters. Physical systems of the same universality class may share the same critical properties, independent of the details of the problem formulation or mathematical model. Universality thus allows us to assert that, near the critical point, \say{the model describes the nature of a real system exactly}\supercite{thurner2018}, ensuring the model's physical correctness. Conversely, a model that fails to exhibit the expected critical behavior can be safely dismissed.

SOC is simply absent in traffic models that rely on strong equilibrium assumptions, such as the Lighthill-Whitham-Richards (LWR) \supercite{lighthill1955,richards1956} theory and its multispecies extensions \supercite{benzoni2003,nair2011}, where the system is completely determined by the initial conditions. Even so-called \emph{nonequilibrium} \supercite{zhang1998} traffic models only add a relaxation term to the dynamics, whereas true nonequilibrium processes capable of criticality and phase-transitions are driven, noisy systems  of \emph{active matter} \supercite{vicsek1995, bain2017, schmiedeberg2023}. In these systems, criticality emerges from the interplay between relaxation and stochastic driving terms \supercite{thurner2018}.

The incompatibility of equilibrium theories with recent empirical evidence becomes apparent in the pNEUMA field experiment, where various types of scooters and motorcycles—hereinafter referred to as microvehicles \supercite{OHern2020}—appear with a notable frequency of up to 30\%. Predicting the erratic dynamics of these microvehicles lies beyond the scope of the LWR model, which is widely regarded as the foundation of traffic flow theory. In the absence of a reliable alternative, transportation researchers tend to disregard the micromobility aspect of pNEUMA by excluding all microvehicle data. However, this common practice is methodologically flawed. To understand why, we must introduce the idea of \emph{anticipation}.

Anticipatory models \supercite{ondvrej2010, moussad2011, karamouzas2014, xu2021} involve a spatiotemporal metric of \emph{time to collision} $\Delta t$ that is defined as the anticipated time at which pairs of non-omniscient, visually obstructed agents, approximated here as inscribed, hard elliptical discs \supercite{zheng2007}, will collide if they keep converging at their current velocities (Fig. \ref{fig:fig1}a-b). $\Delta t$ is particularly useful because theoretical equilibrium conditions are attained when $\Delta t \to \infty$. However, Fig. \ref{fig:fig1}b shows that, even if we only plot car data, $\Delta t$ has a textbook \emph{power-law objective distribution}\supercite{thurner2018} $q\propto\Delta t^{-\mu}$ that is statistically sound \supercite{stumpf2012}. To the best of our knowledge, this power-law is reported here for the first time, indicating that equilibrium conditions occur as rare events drawn from a heavy-tailed distribution. It also empirically verifies that multispecies traffic is an intrinsically far-from-equilibrium system.

Before diving into the results, a brief note on the overall methodology is provided to guide the reader. Instead of developing a theory-based model, simulating it, and then conducting a controlled experiment in the real world to validate a preconceived theory, we do the opposite: we propose a theory-agnostic, \emph{heuristic} \supercite{gigerenzer2008} model calibrated on field data from an uncontrolled experiment, and then identify a theory that explains our results. This approach works for two reasons: Firstly, if the simulation fits the real data as well as the theoretical expectation, then, by causal association, the identified theory should offer a satisfactory explanation for the underlying data-generating natural process. Secondly, some variables might not be readily observable in reality, but are accessible to the analyst in a simulated environment. 

\newpage

\section*{Results}

In line with our empirical findings, we propose in this communication an intrinsically nonequilibrium model for multispecies traffic, which consists of two pillars: a steering and a velocity module. By coupling two simple nonlinearities in the respective modules, we obtain a novel mechanism for 1D directed percolation in continuous space and time. This mechanism can also be interpreted from a statistical point of view in terms of a \emph{sample space reducing} (SSR)\supercite{corominas2015,corominas2017} process.

Using custom simulation software, we model a straight road with two lanes, each of standard width $W=$ 3.6 m and length $L=$ 90 m, with periodic boundary conditions. For the initial conditions, we employ Poisson disk sampling \supercite{bridson2007}, a fast dart-throwing algorithm, to generate feasible, randomized spatial distributions of microvehicles for any given linear car density (Supplementary Fig. 1). Samples are then drawn \emph{without replacement} from this feasible set. Our results are based on ensemble averages from 256 independent simulation runs. Each simulation begins with all agents at rest and concludes after a duration of $\delta t =$ 180 s, with only the last $\delta t^*=60$ s reported, as they are sufficiently far from the transient phase. In stark contrast to traditional equilibrium methods, we do not require an initial perturbation event, since the system is constantly driven far from equilibrium (\href{https://github.com/GAnagno/multispecies-traffic/blob/main/movie-s2.gif}{see Supplementary Movie 2}).

\subsection*{Steering with dynamic view range} Navigating an out-of-equilibrium crowd is a tough task that involves decision-making under uncertainty and incomplete information. In this situation, agents adapt their steering behavior by  adjusting their field of view $\gamma_{\max}$ (Fig. \ref{fig:fig1}c-d) in a dynamic, nonlinear manner, based on their instantaneous speed $v$: \begin{equation}\label{eq:1} 
\log\gamma_{\max}(v) =  b v + c,
\end{equation} where the parameters $b$ and $c$ are obtained by means of \emph{quantile regression} \supercite{koenker2001} (see Supplementary Table 1). Equation (\ref{eq:1}) only provides a dynamic upper bound, while the actual angle $\gamma<\gamma_{\max}$ for each maneuver is the result of a stochastic process, as explained below.

From the angle $\gamma_{\max}$, we define a dynamic view range as the discrete choice set $\alpha \in \left\{\alpha^-,\alpha^+\right\}$, where $\alpha$ has a half-degree resolution in the reference frame of the vehicle, such that $\alpha^+-\alpha^-=2\gamma_{\max}$. Next, the time to collision at the \emph{desired velocity} $v_0$ is calculated for each angle $\alpha$, denoted $\Delta t(\alpha)$ in equation (\ref{eq:2}) \supercite{moussad2011}. Given a horizon distance $d_{\max}$ that depends on the system size, we specify the distance to collision $f(\alpha)$: \begin{equation}\label{eq:2}
f(\alpha)\coloneq\min\{d_{\max}, v_0 \Delta t(\alpha)\},
\end{equation} which is always defined, even if $\Delta t$ is not identified, and is a measure of \emph{dynamic porosity}. Comparing the situation in Fig. \ref{fig:fig1}a with its perception in Fig. \ref{fig:fig1}d, we observe that $f(\alpha)$ can behave counterintuitively, contradicting the static \supercite{nair2011} understanding of porosity. The objective of the dynamic porous heuristic is to probe for an optimal collision-free direction $\alpha^*$ that also minimizes the spatiotemporal detour $d(\alpha)$ from a target direction $\alpha_0$ (square form used for efficiency):  \begin{equation}\label{eq:3}
d(\alpha)^2 = d_{\max}^2 +f(\alpha)^2 - 2 d_{\max}f(\alpha)\cos(\alpha_0-\alpha).
\end{equation} A \emph{random choice} is made if multiple solutions for $\alpha^* = \mathrm{argmin} {d(\alpha)^2}$ exist. In contrast to models of homogeneous crowds \supercite{moussad2011, guo2021}, where $a_0$ points to a fixed egress location (e.g., a gate), we propose a dynamic targeting mechanism. Because in multispecies traffic larger vehicles act as moving obstacles, microvehicles propagate towards the direction of least resistance or higher porosity. Defining $g \coloneqq \max\{f(\alpha^-),f(\alpha^+)\}$, we aim to find $\alpha_0$, such that $f(\alpha)$ is maximized: \begin{equation}\label{eq:4} 
     a_0=\begin{cases}
    0,&g=d_{\max}\\
\mathrm{argmax} {f(\alpha)},&g<d_{\max}.\\
    \end{cases}
\end{equation} If $\Delta t$ is undefined at the limits of the view range, the agent maintains $\alpha_0=0$. Steering is then governed by the following ordinary differential equation (ODE): \begin{equation}\label{eq:5} \dot \alpha = \uptau^{-1}\left[\alpha^*-\alpha\right],\end{equation} where $\uptau$ is a constant time-relaxation. For simplicity, we assume no lateral dynamics for cars in this model.

At this point it should be clear that a dynamic view range is equivalent to an SSR process in the sense that each decision results to a nonlinear adjustment of the sample space, much like rolling a shape-shifting die \supercite{corominas2015}. The view range can also be understood as a measure of noise \supercite{bak1987, vicsek1995} or a (directed) percolation \supercite{stauffer1994, hinrichsen2000} probability.

\subsection*{Heterogeneous optimal velocity model}

\begin{figure}%
\centering
\includegraphics[width=\linewidth]{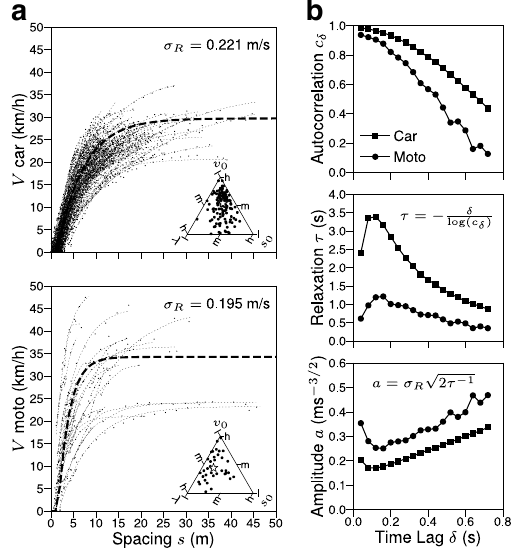}
\caption{Model estimation considering distributed deterministic parameters and a stochastic process. (\textbf{\fontfamily{phv}\selectfont
a}) Speed-spacing responses fitted on pNEUMA data (Supplementary Data 3). Each curve corresponds to strictly 1 human subject, but each individual may engage in different dynamics. We consider in total 29208 driving observations corresponding to $C_{\text{car}}=$ 135 curves, and 2976 riding observations corresponding to $C_{\text{moto}}=$ 38 curves. The ratio ${C_{\text{moto}}}/{(C_{\text{moto}}+{C_{\text{car}})}}$ is representative of the split in the real data with 20-25\% motorcycles. Ternary plots show the curves of each species as points in 3D parameter space using barycentric coordinates. (\textbf{\fontfamily{phv}\selectfont
b}) Model error is captured by a stochastic process. First, the autocorrelation functions $c_{\delta}^{\text{car}}$ and $c_{\delta}^{\text{moto}}$ of the residuals are calculated for several increasing time lags $\delta$ in steps of 0.04 s, that is identical to the time resolution of the dataset. We can then compute the noise relaxation time $\tau = - \delta / \log(c_{\delta})$ and the noise amplitude $a = \sigma_R \sqrt{2 \tau^{-1}},$ where $\sigma_R$ is the standard deviation of the residuals from the nonlinear least squares estimation. A lag $\delta=$ 0.12 s maximizes the relaxation time and minimizes noise amplitude, so we chose it as the time step in numerical simulations.}
\label{fig:fig2}
\end{figure}

In this section we develop a novel optimal velocity (OV) \supercite{bando1995} model that is \emph{heterogeneous, stochastic and nonlinear}. An OV is a phenomenological mapping of the spacing $s$ to an optimal velocity $V(s)$, where $V$ must satisfy general conditions of continuity and monotonicity. We start be reviving a functional form for $V$ that was first introduced as part of Newell's Nonlinear Model (NNM) \supercite{newell1961, daganzo2001}. As Newell writes, \say{no motivation for this choice is proposed other than the claim that it has approximately the correct shape and is reasonably simple}. Although originally proposed for homogeneous car traffic, this nonlinearity predicts microvehicle dynamics surprisingly well (Fig. \ref{fig:fig2}a).

The NNM has 3 shape parameters plus an explicit reaction time, which we substitute by an implicit speed adaptation time. The shape parameters are the jam spacing $s_0$, the slope at the jam spacing $\lambda=V'(s_0)$ and a desired velocity $v_0$. Accordingly, $V:s \mapsto V(s)$ has the following formal specification: \begin{equation} \label{eq:6}
V(s) \coloneq \max\{0, \ v_0 -v_0\exp\left[-\lambda v_0^{-1}(s-s_0)] \right\}.
\end{equation}

Heterogeneity arises when each agent $n$ receive their own response curve $V_n$. As a consequence, the 3 deterministic parameters are distributed and model calibration becomes a two-staged problem. Firstly, the parameters for each driver/rider are estimated by nonlinear least squares fits (Fig. \ref{fig:fig2}a). Secondly, positive continuous distributions are estimated from the obtained populations of empirical parameters by standard maximum likelihood estimators, assuming independence (Supplementary Table 3-4). This is a reasonable assumption following from the fact that the chosen shape parameters are intended to capture completely different behavioral qualities and should be ideally uncorrelated.

Last but not least, temporal stochasticity is a key feature of nonequilibrium traffic. We assume that each $V_n$ is imperfect and subject to perturbations from an error source $\varepsilon_n$ modeled as an Ornstein-Uhlenbeck (OU) stochastic process, resulting in: \begin{equation} \label{eq:7}
    \begin{cases}
    \dot \varepsilon_n = -\tau^{-1}\varepsilon_n+a \xi_n,\\
    \dot v_n = \uptau_n^{-1}\left[w\left(V_n+\varepsilon_n\right)-v_n\right],\\
    \dot x_n = v_n,
    \end{cases}
\end{equation} where $\tau$ is the noise relaxation time  and $a$ is the noise amplitude \supercite{tordeux2016}, estimated by analyzing the stochasticity in our data as explained in Fig. \ref{fig:fig2}b. In order to overcome a well known limitation of the OV model, we also include the weighting factor $w(\Delta t)$, which introduces a dependence on time to collision, making the model anticipatory and \emph{collision-free} even under strong nonlinear perturbations: \begin{equation} \label{eq:8 }
    w(\Delta t) = \frac{1}{2}\left[1+\tanh{A\left(-\frac{1}{\Delta t}+B\right)}\right],
\end{equation} where the parameters $A,B$ are easily derived from first principles \supercite{mammar2005}, effectively switching to deceleration when $\Delta t<1.5$ s. Finally, the speed adaptation time $\uptau_n$ is derived by stability considerations (Methods).

\subsection*{Effect of traffic mixture on the capacity}
\begin{figure*}
\centering
\includegraphics[width=\linewidth]{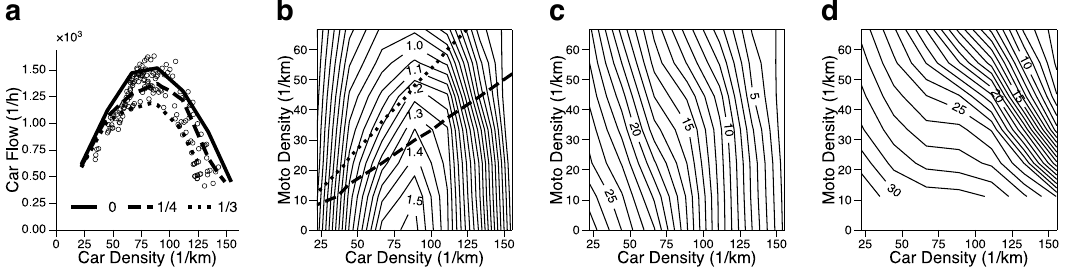}
\caption{Validation of our nonequilibrium model and numerical simulations under diverse traffic conditions. (\textbf{\fontfamily{phv}\selectfont
a,b}) Average car flow per lane. The ensemble average flow $\langle Q\rangle$ is calculated as normalized vehicle kilometers traveled (VKT): $\langle Q\rangle = \frac{\langle\text{VKT}\rangle}{
         (\delta t^* / 3600) (2L / 1000)}$, where $\delta t^*$ is the effective duration of the simulation and $2L$ is the total lane length. Lines represent simulation results for different motorcycle shares, whereas dots depict quasi-stationary experimental measurements (Supplementary Data 4). (\textbf{\fontfamily{phv}\selectfont
c}) Average car space-mean speed. (\textbf{\fontfamily{phv}\selectfont
d}) Average motorcycle space-mean speed. We observe that micovehicles are considerably faster than cars.}
         \label{fig:fig3}
\end{figure*}

The effect of traffic mixture on the road capacity is evaluated on a finite grid of vehicle concentrations. The capacity-related externalities of micromobility are evident in Fig. \ref{fig:fig3}a-b, where we show a diagram of car flow per lane for different motorcycle shares, which is also in agreement with experimental data. Furthermore, an effective lane capacity of 1600 veh/h agrees with the values reported in the literature for similar infrastructures \supercite{wu2011}. We observe that for a small share of motorcycles, the detrimental effect on car flow is rather negligible and maybe this explains why microvehicles are often not taken into account when studying urban traffic. However, as the motorcycle concentration increases further, the car flow drops drastically in a strongly nonlinear fashion. This is a rather surprising result that contradicts the only known previous numerical study (not validated) \supercite{lan2010}, where a linear reduction in car flow was found. It also appears that the system attains maximum flow at a higher car density than previously reported. Finally, surface plots of space-mean speed as monotonically decreasing function of multispecies concentration are shown, for both species (Fig. \ref{fig:fig3}c-d). Interestingly, for a broad range of traffic mixtures, motorcycle speed becomes insensitive to variations in the car density, a finding which suggests spontaneous separation of flows and merits further, quantitative investigation. 

\subsection*{Multispecies traffic as DP} While one can argue that the above aggregate traffic variables can be potentially reproduced with a more coarse-grained model, the advantage of our approach is that it also enables the study of nonequilibrium critical phenomena. In this section, we examine the hypothesis that multispecies traffic is an instance of DP, the \say{most important class of nonequilibrium processes, which display a nontrivial phase transition} \supercite{hinrichsen2000}. Phase transitions involve a control parameter and a state variable or order parameter \supercite{stanley1987}. The control can be for example noise or temperature. In general there is a non-universal critical threshold $p_c$ separating a disordered, subcritical phase from an ordered, supercritical phase and also a universal critical exponent that characterizes the transition. DP in particular is characterized by 3 critical exponents: $\beta$, $\nu_\perp$ and $\nu_\parallel$. These constants cannot be found analytically, but are known up to an arbitrary numerical accuracy and correspond to 3 variables of state: density, spatial and temporal correlation lengths \supercite{hinrichsen2000}. To support the case for DP in multispecies urban traffic, we adopt an \emph{off-lattice formalism} \supercite{vicsek1995} that operates on a moving substrate, rather than the conventional DP on static lattices. This methodological difference should have no effect on the universality of the critical exponents. 

Starting with the control parameter, we define the mean view range of all microvehicles at each time step as the temperature $T$, which is a spatial average: \begin{equation}\label{9}
T \coloneq \frac{1}{2}\text{nint}\frac{1}{N} \sum_{n=1}^{N}\alpha_n^+-\alpha_n^-,
\end{equation} where the time dependence is omitted for notational ease and the nearest integer operator is discretizing the view angles at half degree precision. The sum runs over a snapshot of the lane-free population $N$.

Unlike models of homogeneous systems, the order parameter here must be a relative measure to account for the moving substrate. For each species $i$, we define the instantaneous order parameter $\Phi_i$ :\begin{equation}\label{eq:10}
    \Phi_{i} \coloneq \frac{1}{N_{i}}  \left\Vert\sum_{n=1}^{N_{i}}\frac{\bm v_n}{V_{n}(d_{\max})}\right\Vert,    \ i \in \left\{\text{car}, \text{moto}\right\},
    \end{equation} where $\bm v_n$ denotes the velocity vector of agent $n$ and $V_{n}(d_{\max})$ the respective maximum attainable speed in the simulator. We then compute the difference $\Delta \Phi= \Phi_{\text{moto}} - \Phi_{\text{car}}$ as a measure of percolation. When $\Delta\Phi>0$, motorcycles are on average faster and may percolate through the cars. However, when $\Delta\Phi<0$, cars are more efficient because motorcycles are over-maneuvering (substrate is faster).

\begin{figure*}
\centering
\includegraphics[width=\linewidth]{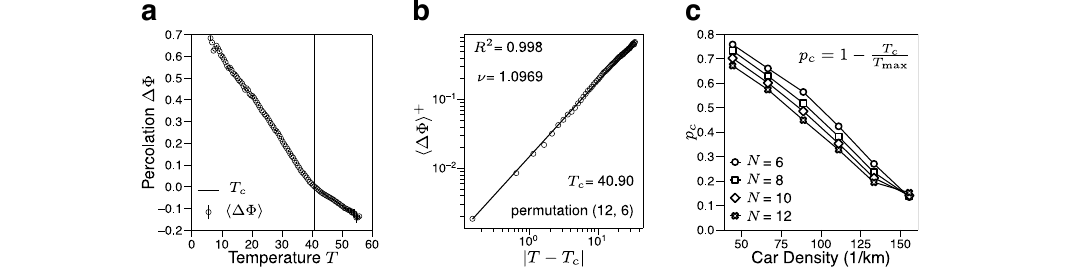}
\caption{Nonequilibrium phase transition in multispecies traffic. (\textbf{\fontfamily{phv}\selectfont
a}) Spatial ensemble averages of the percolation index $\Delta \Phi$ over all simulation runs plotted as a function of $T$ for a specific traffic permutation with 12 cars/lane and $N=6$ motorcycles. The error bars, which have been obtained by bootstrapping, are barely visible and smaller than the marker size. A sharp discontinuity occurs at the critical temperature $T_c$ despite the relatively small system size with $L=90$ m. (\textbf{\fontfamily{phv}\selectfont
b}) For the same traffic permutation, $\langle \Delta \Phi \rangle_T^+$ follows an exact power-law having an $R^2\approx1$ and a critical exponent $\nu\approx\nu_\perp$. (\textbf{\fontfamily{phv}\selectfont
c}) Non-universal normalized percolation thresholds $p_c$ is found by substituting $T_{\max}=$ 56 (see Fig. \ref{fig:fig1}c). Each marker corresponds to a statistically significant power-law fit with an average $\overline{R^2}=0.993$ and a standard deviation $\sigma_{\nu}=$\num{1.4e-5}.}
\label{fig:fig4}
\end{figure*}

In order to measure the phase transition, we fix the density and aggregate the simulated data by their temperature $T$. For each $T$, the ensemble average $\langle \Delta \Phi \rangle_T$ is computed over all realizations and steps of the random process. Around a critical temperature $T_c$, that is not universal and depends on the density, $\langle \Delta \Phi \rangle_T \to 0$ and a nonequilibrium phase transition occurs, as shown in the example of Fig. \ref{fig:fig4}a. For the rest of the analysis, we ignore $\langle \Delta \Phi \rangle_T < 0$ and focus solely on the supercritical case of positive percolation $\langle \Delta \Phi \rangle_T^+$. The inverse quantity $1/{\langle \Delta \Phi \rangle_T^+}$ would then be analogous to a spatial correlation length that diverges at criticality, where the multispecies system does not have a characteristic scale. This phase transition is governed by a power-law (Fig. \ref{fig:fig4}b):\begin{equation}\label{eq:11}
    \langle \Delta \Phi \rangle^+_T \sim \left|T-T_c\right|^\nu,    \ T<T_c ,    
    \end{equation} where $\nu$ is the universal critical exponent. Since DP is a non-integrable process with both path and history dependence, it cannot be solved exactly and no analytical formula exists for the critical threshold \supercite{hinrichsen2000}. This implies that equation (\ref{eq:11}) contains two unknowns: $T_c$ and $\nu$. For each permutation, we fit a distinct power-law by also imposing the optimality condition $T_c = \mathrm{argmin} {\left|\nu-\nu_\perp\right|}$, where $\nu_\perp$ is the theoretically expected value of the spatial correlation exponent for 1D DP, as reported in \supercite{hinrichsen2000}. Rather than maximizing the goodness of fit, our objective is to minimize the deviation from the theoretical value, thus producing consistent estimates of the critical thresholds.
    
    Our 1D DP hypothesis is unambiguously confirmed with an average $\overline{R^2}=0.993$, for the statistically significant density permutations shown in Fig. \ref{fig:fig4}c. These results are presented in greater detail in the appendix (see Supplementary Fig. 2). Accordingly, the normalized critical threshold $p_c$ is clearly non-universal and depends largely on the car density per lane, which indicates the level of traffic congestion. It also appears that the number of microvehicles $N$ has negligible effect on criticality. A lower percolation threshold with higher car density implies that there is a greater chance that even a random arrangement of motorcycles will successfully percolate through the traffic, a finding that aligns with intuition, given the considerable agility of microvehicles.

    It is quite remarkable that the complex dynamics of multispecies urban traffic can in fact be reduced to 1 irrational number (Fig. \ref{fig:fig4}b), which is universal and common among all the other porous flows of the same dimensionality, at a macroscopic level. This suggests that, near the thermodynamic limit, heterogeneous, multispecies traffic behaves in the same way as water percolating through a porous medium.

\section*{Discussion}
In this work we revisit traffic flow theory from the standpoint of complex systems, by first showing that urban traffic is a far-from-equilibrium system and then demonstrating with a nonequilibirum model that multispecies urban traffic has indeed porous characteristics and belongs to the DP universality class. Our result opens a new research avenue for the application of percolation theory to transportation, beyond the scope of existing network models \supercite{li2014, olmos2018, saberi2020, zeng2020, ambhl2023}. The nonequilibrium model also accounts for the large scatter observed in recent field data by considering the coexistence and competition between lane-based and lane-free flows. Our open simulation framework can be used to evaluate future scenarios and policies, such as the anticipated increase in lane-free traffic or the performance of connected and automated vehicles in challenging multispecies environments. 

It would be interesting and valuable to explore if the universality principles outlined above hold true for other types of heterogeneous crowds, such as a mixed population of pedestrians and e-scooters, or even non-human crowd mixtures. We note that a more comprehensive treatment of the subject should include all 3 exponents that uniquely characterize 1D DP, as well as an investigation of the $\langle \Delta \Phi \rangle_T < 0$ case. Another promising research direction involves the development of a lattice-based, discrete version of our model, which may offer faster computation and be more conducive to theoretical analysis. Finally, we wish to relax the assumption that cars have no lateral dynamics as it is known that lane-changes contribute to traffic congestion and instabilities \supercite{laval2006}. Therefore, a lane-changing \supercite{kesting2007} extension tailored to multispecies traffic is an important next step.

\section*{Methods}

\subsection*{Maneuver detection} The real data shown in Fig. \ref{fig:fig1}c are produced by our maneuver detection algorithm. This method requires 3 inputs: vehicle positions ${(x,y)}$, vehicle azimuth $\theta_{\text{veh}}$ and road azimuth $\theta_0$. Then, we proceed to define the instantaneous relative angle $\theta=\theta_{\text{veh}}-\theta_0$ and the curvature $\kappa=\dv{\theta}{r}$ with respect to the arch traveled ${r}$. Trajectory points with ${\kappa = 0}$ are essentially inflection points, where the curvature changes sign or, in terms of driving/riding, a steering event occurs. Steering events occur regularly in cruising but become sparse during maneuvering. By detecting such sparsity, we can reliably detect and characterize any maneuver.

\subsection*{Leader identification} Unlike constant-speed simulations, such as the ones produced with Vicsek's model, leader identification is relevant in dynamic traffic models. More specifically, leading agents in 2D are determined by a heuristic known as the overlap criterion, which is explained in \supercite{xu2021} and references therein. This is simply a stripe, as shown in Fig. \ref{fig:fig1}a, of exactly the same width as the ego vehicle and infinite length. Taking into account the size of our simulator and the fact that periodic boundary conditions are in place, we set the stripe length equal to the horizon distance $d_{\max}=40$ m. Unfortunately, this criterion neglects lateral frictions that interrupt the pure follow-the-leader behavior, a phenomenon that has been empirically observed in the pNEUMA experiment. Consequently, we enlarge the stripe width by a factor $f=$ 1.5 that permits the detection of laterally adjacent agents. Our adapted overlap criterion introduces asymmetric effects as it depends on vehicle size, which varies considerably in traffic mixtures. Including lateral friction prevents unrealistic crashes between microvehicles and also generates realistic nonlinear perturbations in the car platoons, which have an impact on the capacity.

\subsection*{Stability considerations}
Consider $N$ vehicles from the same or different species. The equilibrium spacing $s_n^*$ of each vehicle $n$ is found by solving for the equilibrium velocity $v^*$: \begin{equation}\label{eq:12}
  \sum_{n=1}^N s_n^*=\sum_{n=1}^N V_n^{-1}(v^*)=2L-\sum_{n=1}^N \ell_n ,
\end{equation} subject to $0<v^*<\inf_{n}v_n^0$. $2L$ is the total road length, including both lanes, and $\ell_n$ is the vehicle length. Then from the strict stability criterion for heterogeneous traffic which is known as combination stability \supercite{yang2014}, we obtain the implicit speed adaptation time $\uptau_n$ for each agent $n$: \begin{equation}\label{eq:13}
  \uptau_n = \max\left\{\left[1+\cos{\frac{2\ \pi}{N}}\right] V_n'(s_n^*),\lambda_n\right\}^{-1},
\end{equation}

subject to $s_n^*>s_n^0$. The term in square brackets approaches 2 as $N\to\infty$ resulting in more precise stability bounds for finite rings \supercite{huijberts2002}.  Notice that the stochastic component of the model does not impact the stability because it only captures imperfections in the optimal velocity function as opposed to driver control or acceleration errors \supercite{laval2010,ngoduy2019}.

\subsection*{Numerical scheme}
For the numerical computation of $\Delta t$ we use a Newton-Raphson iteration with tolerance of \num{e-4} s. The solution to the stochastic differential equations follows the Euler-Maruyama integration scheme \supercite{tordeux2016, ngoduy2019} with a time step $\delta=$ 0.12 s: \begin{equation} \label{14}
    \begin{cases}
    \varepsilon_n(t+\delta) = (1-\delta\tau^{-1} )\varepsilon_n(t) + a \sqrt{\delta} \eta_n,\\
    v_n(t+\delta) = v_n(t) + \delta\frac{ w\left[V_n +\varepsilon_n(t+\delta)\right] -v_n(t)}{ \uptau_n},\\
    x_n(t+\delta) = x_n(t) + \delta v_n(t+\delta),
    \end{cases}    
\end{equation}

where each independent $\eta_n \stackrel{iid}{\sim} \mathcal{N}(0,\,1)$. In terms of operational navigation, agents steer smoothly with a global relaxation time $\uptau=$ 0.6 s, a value within the range reported in several other studies \supercite{moussad2011,guo2021, xu2021}: \begin{equation}\label{eq:15} 
\alpha_n(t+\delta) = \alpha_n(t) +\delta\frac{\alpha_n^{\text{des}}(t)-\alpha_n(t)}{\uptau}.
\end{equation}

\subsection*{Data availability}
We analyzed extensive field data, which is available at \url{https://open-traffic.epfl.ch}. The data sample used in this paper was collected on October 24, 2018, between 9:30 and 10:00, using drone D8. The sample was preprocessed using anomaly detection and data smoothing techniques specifically designed for traffic monitoring with a swarm of drones \supercite{kim2023}.

\subsection*{Code availability}
The custom simulation software for replicating the methods and results outlined in the paper, along with the corresponding documentation, is publicly available on GitHub at: \url{https://github.com/EPFL-ENAC/LUTS-pneuma-simulator}.

\printbibliography
\section*{Acknowledgements}
Our paper was partially funded by the Swiss National Science Foundation (SNSF) grant 200021\_188590 “pNEUMA: On the new era of urban traffic models with massive empirical data from aerial footage”.
\section*{Author contributions}
GA and NG designed the research. GA performed the numerical simulations. GA and NG developed the theory, discussed the results and wrote the paper.
\end{document}